\documentclass[fleqn,10pt]{wlscirep}
\usepackage[utf8]{inputenc}
\usepackage[T1]{fontenc}
\usepackage{float}

\newcommand{\bell}{\boldsymbol{\ell}}

\newcommand{\bog}{\boldsymbol{g}}

\newcommand{\bSigma}{\boldsymbol{\Sigma}}

\newcommand{\bPsi}{\boldsymbol{\Psi}}

\newcommand{\bmu}{\boldsymbol{\mu}}

\newcommand{\bs}{\boldsymbol{s}}

\newcommand{\bx}{\boldsymbol{x}}
\newcommand{\bX}{\boldsymbol{X}}
\newcommand{\by}{\boldsymbol{y}}
\newcommand{\bY}{\boldsymbol{Y}}

\newcommand{\bzero}{\boldsymbol{0}}

\newcommand {\bN}{\mathcal{N}}

\title{Probabilistic learning to perform pre-onset individualised prediction of disease severity: application to Veno Occlusive Disease}

\author[1,*]{Dalia Chakrabarty}
\author[2]{Kane Warrior}
\author[3]{Chuqiao Zhang}
\author[4]{Akash Bhojgaria}
\author[5]{Joydeep Chakrabartty}

\affil[1, 2, 3]{University of York, Department of Mathematics, York, YO10 5DD, U.K.}
\affil[4,5]{Narayana Superspecialty Hospital, Kolkata 711103, India}

\affil[*]{dalia.chakrabarty@york.ac.uk}

\begin{abstract}
We advance a new probabilistic supervised learning approach that permits reliable, automated, and early individualised prediction of the severity with which a disease will develop in a prospective patient. The prediction capacity is illustrated via the pre-transplant prediction of the score of severity of Veno Occlusive Disease (or VOD) in the digital twin (DT) of the considered prospective patient, where this score parametrises the severity with which VOD will develop in this patient, after they undergo their Bone Marrow Transplant. The learning of the relationship between the pre-transplant variables, and a severity score variable is undertaken by modelling this relationship as a (random) function that is treated as a sample function of an adequately-chosen stochastic process. The parameters of this underlying process are learnt using a training dataset that is generated using the real-time evolution of retrospective patients in a cohort, with this training dataset subsequently augmented in size by a probabilistic inverse learning of the score of prospective patients. The augmented training set, then permits the learning of the function that capacitates - at the pre-transplant stage - automated prediction of the score of the severity of VOD that characterises the DT of a physical patient in their unique pre-transplant state. This score is subsequently fed back to the real prospective patient as the severity with which VOD will develop in them, after this patient undergoes their transplant. Such a score then permits the treating Haematologist-Oncologists to decide on the treatment regimen, which in this illustration reduces to deciding on treating the patient with Defibrotide. An AI facility is developed to undertake such automated prediction, with the physician inputting the data on the pre-transplant state that characterises the DT of the prospective patient under consideration.
\end{abstract}
\begin{document}

\flushbottom
\maketitle

\thispagestyle{empty}

\section*{Introduction}
\label{sec:intro}
Inhomogeneities in the susceptibility to a disease within patient cohorts is an undeniable truth, triggered by unique genotypic and phenotypic details of patients. The response to this reality is individualised medicine \cite[amongst others]{eu, nhs, cruk, blackstone, hetero_tum} that promises better diagnosis and management of the disease. In this paper, we discuss a methodology for achieving reliable individualised prediction at the pre-onset stage, of the severity with which a disease will develop later in a prospective patient. Thus, the undertaken prediction is firstly individualised, since the severity with which the disease will develop in the patient is driven by the pre-onset state of the individual patient. The prediction is also early, and this is of significance in the treatment of multiple diseases, (especially diseases that are potentially terminal), with a correlation established between early diagnosis, and survivability \cite[out of many other references]{nhs_early, cruk_early, crucial_early}.

We illustrate the predictive capacity of our new approach on the disease called Veno Occlusive Disease (or VOD) that often develops in recipients of haemopoietic stem cell transplantation, which we shall refer to below as ``Bone Marrow Transplants'', and abbreviate throughout as BMT. VOD, (which is also referred to as Sinusoidal Obstruction Syndrome in the literature), results from the chemo-irradiation therapy relevant to BMT. It typically develops in a BMT-recipient within a few weeks of the BMT, and VOD can be life-threatening, with the prevalence of severe VOD noted in be about 33$\%$ of cases, of which, it can be terminal in about 80$\%$ cases \cite{nhs_vod}. Severe VOD typically calls for prolonged hospital stay, ``including the need for intensive care support'' \cite{nhs_vod}. Defibrotide is considered the only effective treatment of VOD in its severe manifestation, with response rates of 24$\%$-49$\%$. It is noted to significantly improve mortality compared to historical interventions, and to shorten hospital stay.

Given this, Defibrotide is used widely to treat VOD, with NHS UK
suggesting that patients be started on Defibrotide four times a day,
from the time that VOD symptoms show up in the patient, for 21 days or till the VOD symptoms ebb
\cite{nhs_vod}. However, VOD symptoms might be conflated with the
symptoms of other underlying diseases \cite{bonafazi, angus, yoon}, and importantly, VOD
survivability is noted in the literature as strongly correlated with
early intervention \cite[amongst others]{mehra, richardson, bonafazi}. Given this, we seek a method to predict
a score of the severity with which VOD will develop in a
BMT-recipient, but our aim is to seek such a score at the pre-transplant stage,
i.e. early. It is the set of pre-transplant variables that informs
on the individualised susceptibility to VOD severity, that
characterises the prospective BMT-recipient.

Thus, we use
such pre-transplant variable information to model a prospective
patient as their digital twin (DT), and predict the severity score for this
DT, in order to capacitate the decision-making towards the treatment regimen that is to be followed, for this considered (real) prospective BMT-recipient.
We recall that a DT is a virtual copy of a real patient, and is formulated by collating the patient’s health data, which in our consideration, is the data on the pre-transplant state of the considered prospective patient. Such usage of DTs of real patients is an emerging trend within healthcare today \cite{scoping, hospital_times, meijer, sun_he}.

In our work, a prospective patient whose DT is assigned a predicted mean score $\geq$ 0.11, is interpreted to suggest that this patient will develop VOD, post-BMT, as is discussed below. A pre-transplant prediction of a mean score of $\gtrsim 0.45$ in the DT is interpreted as saying that the BMT will induce a moderate-to-high severity of VOD in the patient. When they can, the treating Haematologist-Oncologists decide to treat such a patient with Defibrotide from Day~-2, (i.e. two days before the transplant), till about Day~20, (i.e. the 20-th after the BMT). In fact, our prediction follows the learning between the severity score variable and the set of pre-transplant variables, which in itself is rendered possible using a training dataset comprising the severity scores of each retrospective patient in a cohort, where we know the values of the pre-transplant variables of the respective patient. 

Indeed, the learning of the relationship between the vector of pre-transplant variables and the VOD severity score variable, required such a training dataset. However, contrary to the demands of this proposed learning, we did not possess information on values of the severity score variable, at known values of the pre-transplant variables of any patient. Hence, said learning appeared to be in jeopardy. To address this problem, we had learnt the relative severity score of each patient in a retrospective cohort, i.e. a cohort in which each patient had already undergone their BMT. We reported the said score learning in \cite{plos}. We had used the time series data on multiple physiological parameters that were recorded in the patient charts - from a few days before the BMT, to a time point after the same - to learn random graphs of each patient in the cohort. Said time series data were of disparate lengths, since some patients expired early, given the terminal nature of their illness. Acknowledging such challenges of the available data, we invoked a statistical distance between the probability distribution of the random graph variable learnt given the time series data of a patient, and that of an arbitrarily-earmarked ``reference'' patient, such that this distance allowed us to compute the relative score of the considered patient in the cohort. Since the pre-transplant variables of each patient in this cohort were known, once the relative severity scores were learnt, the originally-absent training data was made available, using which we had learnt the relationship between these two variables as reported by \cite{plos}. We augment this existing training dataset in this work.

We learn the aforesaid relationship, by treating it as a function that is an unknown. In Bayesian statistics, anything that is an unknown, is a random variable, that we can ascribe a probability distribution to. So we model the sought random function as a sample function of an adequately chosen stochastic process. Such a choice is motivated under the ``Background'' section, along with a discussion of the advantages of this learning strategy. One distinct advantage is that we are able to perform the pre-transplant prediction of the mean severity score - and the uncertainties in it - of a new or prospective patient in an automated, reliable and fast way, where pre-transplant variables of this prospective patient are known to us. The overarching aim of this paper is to expound upon a method that allows for such reliable and automated prediction.

\section*{Background}
\label{sec:adv}

\subsection*{Advantages of our method}

It is to be noted that our method is {\it{not}} a machine learning technique; instead, we learn the relationship between a severity score variable, and the vector-valued variable comprising the pre-transplant parameters, as a random function (i.e. a function-valued random variable) that we need to learn, given a training set. This training set comprises pairs of values of the vector of pre-transplant variables of a retrospective patient (who has undergone their BMT), and the score of the severity with which VOD has developed in this patient. We treat the sought random function as a sample function drawn from an adequately-chosen stochastic process \cite{plos}. This model is motivated by the fact that a stochastic process gives the probability distribution of a random function. The choice of the generative stochastic process is motivated by the fact that the shape of the sought function should be minimally constrained by the chosen process. Indeed a Gaussian Process (GP) is such that it only imposes a constraint on the joint probability density of a finite number of realisations of a sample function drawn from this process, (constraining this joint to be a multivariate Normal density). It does not impose an explicit constraint on the shape of its sample function. Again, a GP can be easily extended across dimensions, to model a random function, the output of which is a vector-valued variable, while measurement noise can be organically-folded into the learning of the function using a GP. Very usefully, the probability distribution of the function's output realised at a test input, is fully-known, including its mean and variance. Thus we can compute this predicted output, in a closed-form way.

In \cite{plos} we discussed the method used to initiate an effort towards populating the
training dataset that is a requisite for the probabilistic learning mentioned above. As reported in that paper, we learnt
the relative severity score of each patient in a retrospective cohort, such that each patient in this cohort had already undergone
their BMT, where values of the pre-transplant variables of each such patient were known. We learnt this score as a statistical distance
between the probability distribution of the random graph of a patient in this
cohort, and that of an arbitrarily-identified
``reference'' patient in this cohort. Here, the graph of a patient is
learnt using the time series data manifest in their patient chart that
was recorded during a time interval that includes their BMT. Thus, the graph of a given patient's data embodies the correlation between pairs of the patient's diverse
physiological parameters, that are evolving with time during this
monitored time interval. Then the difference between said
correlation structures of two patients, betrays the severity
of VOD that one of these patients is suffering from, relative to the
other. Updating this training set is relevant to the enabling of automated prediction of VOD severity score, as we will discuss in the subsection below, on closed-form prediction.

\subsubsection*{Small datasets suffice for our probabilistic learning}
A machine learning technique fundamentally requires a large training set, since such techniques learn the inter-variable relationship by identifying patterns in the data comprising pairs of the two relevant variables. This however, is starkly different from our approach in which we learn the parameters that define the generative stochastic process that underpins the random function that we model as the sought inter-variable relationship. In other words, we learn parameters of the probability distribution of the considered function-valued random variable - recalling that a stochastic process offers probability distribution on the space of functions. On the other hand, we do not aim to pick up the patterns of the training data, (with said patterns anticipated to be generally nonlinear), that comprises pairs of values of the two variables, one of which is in fact a high-dimensional vector. Consequently, our method requires far less empirical information than machine learning techniques.

The training data generated by \cite{plos} used information on a retrospective cohort comprising 25 patients. Some results from our score learning are recalled below from \cite{plos}.
\begin{itemize}
\item The reference patient in the retrospective cohort is arbitrarily chosen, and they are arbitrarily attributed a VOD severity score of 1.
\item It is on the scale in which this patient has a score of 1, that any other patient's score is interpreted. 
\item By cross-correlating our scores learnt for each patient in this cohort, with the VOD diagnosis performed by physicians after the BMT of each patient, we identified

that all retrospective patients with mean learnt scores $\geq 0.11$, had developed VOD.
\item Also, all patients with mean learnt score $< 0.11$, were diagnosed to have not developed VOD post-BMT.
  \item Some patients with score $\gtrsim$0.8 were known to doctors as those who had suffered from severe VOD. 
\end{itemize}
In light of this finding, we treat 0.11 as the threshold of the mean score, in excess of which, a score predicted at the pre-transplant stage, is interpreted to imply that the prospective patient will develop VOD after their BMT. We also assert that a higher mean score, implies a higher severity of VOD. 

\subsubsection*{Pre-transplant prediction of a continuous-valued score of severity}
Crucially, we offer a value of a continuous-valued score of the severity with which VOD will develop in any prospective patient, where the score is clearly interpreted. Thus, we will be able to interpret the pre-transplant predicted score as whether VOD will at all develop in the prospective patient after their transplant, as well as comment on the intensity of VOD development in said patient. The method permits the prediction of the mean severity score, as well as the inclusion of comprehensive, interpretable and objective uncertainties on the predicted score. Hence our scoring is much more informative than any categorised assessment of severity levels with which VOD will develop in a given patient post-transplant - unlike in \cite{lee2022}.

Additionally, we do not predict the probability with which VOD will develop at low/medium/high levels, but directly answer the question that physicians seek: ``What is the severity with which the disease will develop in the given patient, once the patient is afflicted by this disease?''. Offering the probability for attaining a pre-defined categorised severity level answers a different question: ``What is the probability with which the patient will develop VOD with severity in the $k$-th level, for $k\in\{{\text{high, moderate, low}}\}$?'', or for the categorised level variable to attain values in another set. The latter question is not particularly useful to physicians when deciding on the pre-emptive treatment/prophylaxis course, if the probability predicted by the learning scheme is nearly 0.5. For example, if the prediction is that there is a probability of 0.55 for the prospective patient to develop VOD with high severity, and a probability of 0.45 for them to develop VOD with moderate severity, it is difficult for the physicians to clearly interpret the offered prediction. On the other hand, when we predict at the pre-transplant stage, that after their BMT, the given patient will develop VOD with a mean severity of 0.45, the Haematologist-Oncologists understand this as a moderately severe level of VOD that will appear in this prospective patient, where the doctors are offered the information that a patient will develop VOD post-BMT, only if their predicted mean score is $\geq 0.11$.

\subsection*{Questioning accuracy of training data used in ML attempts}
Again, the training data that is required for machine learning - which subsequently permits prediction of the probability with which a prospective patient will develop VOD at a given categorised level of severity - is based on physicians' assessment of the categorised severity of VOD that a retrospective patient is suffering from, after they have undergone their BMT. However, VOD literature informs us that symptoms of VOD can be conflated as symptoms of other underlying diseases \cite{bonafazi, yoon}. In fact, \cite{bonafazi} say that ``there are other liver diseases that are common after marrow transplantation'', and these could ``mimic'' VOD as well as ``coexist'' with VOD. Hence, it appears risky to use such data to train a classifier that predicts VOD severity in a prospective patient.

\subsection*{Learning the function that outputs the severity score}
As is motivated above, we would like to learn the relationship between the pre-transplant variables and the severity score, as a function $f(\cdot)$, the input to which is a value $\bx$ of the vector $\bX$ of pre-transplant variables of a patient, such that this function outputs the severity score variable $S$ of this patient. We model this function as a random function that is treated as a sample function drawn from a GP. Thus, the exercise of learning the function is reduced to the learning of the parameters of this GP. To summarise, we seek to learn the function $f(\cdot)$ where $S=f(\bx)$. We refer to this model as Model~1.

\subsubsection*{Original training set too small for Model~1 to be useful}
Parameters of a stochastic process are its mean function and its covariance function. We standardise the data on the severity score, as included in the training set we had available in \cite{plos}, and this motivates using a zero-mean GP, leading to the need to learn only the covariance $Cov(S_i,S_j)$ between the output score variable $S_i$ - that is realised at the input that is the value $\bx_i$ of the pre-transplant vector - and the output $S_j$, realised at input $\bx_j$, for all $i,j\in\{1,\ldots,N\}$, where there are $N$ data points in this training dataset, i.e. there are $N$ patients in the retrospective cohort. $N=25$ in the retrospective cohort that was reported in \cite{plos}. Here, the vector $\bX$ comprises $p=30$ pre-transplant variables, as each of its 30 components. Then for any $i,j$ pair, (i.e. for the $i$-th and $j$-th patients), to learn $Cov(S_i, S_j)$, we will need to learn at least $p=30$ number of parameters, that we refer to as $\ell_1,\ell_2,\ldots,\ell_p$. Here, $\ell_k$ is interpreted as the scale length one needs to move along the $k$-th direction in the space of the 30-dimensional input variable $\bX$, from the point $\bx$ to $\bx^{\prime}$, in order for covariance between output variable $S=f(\bx)$ and output $S^{\prime}=f(\bx^{\prime})$ to fall by a stipulated factor, (from the covariance when $\bx=\bx^{\prime}$). This holds for any $k\in\{1,\ldots,p\}$, Thus, in this learning strategy, we will need to learn 30 length scales. However, our training data includes only 25 data points. Then it will be fundamentally impossible to learn these 30 requisite parameters for the learning of the sought function $f(\cdot)$ with this 25-sized training dataset. 

\subsubsection*{Circumventing problem of small-sized training set}
Hence in \cite{plos} we had resorted to learning the random function $\bog(\cdot)$, that takes a value $s$ of the score variable $S$ as its input, such that it outputs the pre-transplant vector variable $\bX = \bog(s)$. We refer to this model as Model~2. Again, we had modelled $\bog(\cdot)$ as a random sample function of a {\it{vector-valued}} GP. Learning its covariance function had then demanded the learning of only a single length scale parameter $\ell$ (minimally) since the input to the sought function $\bog(\cdot)$ is a value of the scalar-valued variable $S$. Additionally we will need to learn the amplitude parameter $A > 0$. So in \cite{plos} we could undertake the learning of $\bog(\cdot)$ using our available training set. 

The advantage of Model~1 is that it allows for the closed-form - and therefore automated - prediction of the (mean and variance) of the output variable (score) that is realised at a test input, namely the pre-transplant vector of a prospective patient. In a GP-based modelling, only the output prediction is closed-form. However, Model~2 outputs the pre-transplant variable vector $\bX$, the value of which, is in fact known for a prospective patient, and the input score value is then not directly predictable, i.e. score prediction is not automated in Model~2. In fact, in \cite{plos}, we reported the learning of the score of seven prospective patients using Markov Chain Monte Carlo, (or MCMC) based inference undertaken on this score, as well as on $\ell$ and $A$.

\subsubsection*{Augmenting size of training set, to capacitate learning with Model~1}
To avail of the automated score prediction that is possible with Model~1, we had to augment the size of the training set, to capacitate the learning of the 30 length scale parameters $\ell_1,\ldots,\ell_{30}$. In order to achieve this, we undertook the MCMC-based inference with Model~2 in eight other prospective patients - besides the seven prospective patients whose learnt scores were reported in \cite{plos} - and learnt the score of the severity with which VOD will develop after their respective BMT. These eight patients whose scores we learnt post-\cite{plos}, were marked by their IDs: ID~34, ID~35, ID~36, ID~37, ID~38, ID~39, ID~40, ID~41. Of these eight patients, we kept aside the scores of two patients to make (forward) predictions later with Model~1; they are the patients with ID~35 and ID~38. Then with the 7+8-2=13 learnt scores, we updated the training set to a size of 25+13 = 38. For each considered patient, the value of the vector of pre-transplant variables is known.

We then define our augmented training set as ${\bf D}_{aug} = \{(\bx_i, s_i)\}_{i=1}^{38}$. Using the training set ${\bf D}_{aug}$ we learn the relationship between the pre-transplant vector variable $\bX$ and the severity score variable $S$, within Model~1, as the scalar-valued random function $f(\cdot)$ with $S = f(\bx)$.

\subsubsection*{Closed-form pre-transplant prediction of severity score of DT of prospective patient}
\label{sec:predatlast}
\begin{figure}[ht]
\centering
\includegraphics[width=.6\linewidth]{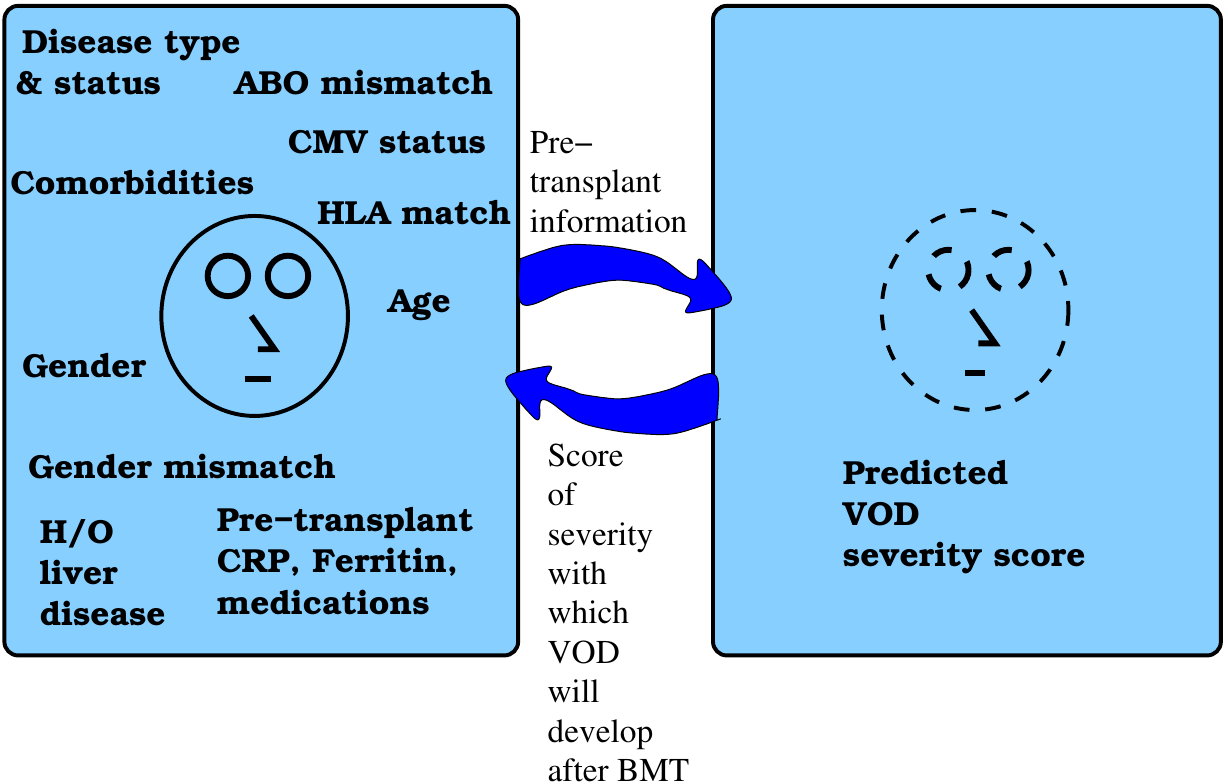}
\caption{The digital twin (on the right) of a prospective patient (on the left), as the representation of the real prospective patient in their pre-transplant state, where this state is characterised by parameters that fall under the headers mentioned in the left panel.}
\label{fig:dt}
\end{figure}

Once we had collated the training dataset required for learning the function that outputs the score, we construct the DT of a prospective patient as the embodiment of the pre-transplant state of this patient. Such a state is parametrised by 30 pre-transplant parameters of a prospective patient, namely:
\begin{itemize}
\item age; gender;
\item disease type - Acute Lymphoblastic Leukaemia, Acute Myleoid Leukaemia, Chronic Myleoid Leukaemia, Aplastic Anaemia, Myelofibrosis, or Others (including Thalassemia); 
\item disease status - relapsed, remission, (such that if each status variable takes a value 0, the staus is interpreted as refractory);
\item comorbidities - hepatic dysfunction, pulmonary dysfunction, cardiac dysfunction, diabetes, hypertension;
\item pre-transplant Ferritin;
\item pre-transplant CRP;
\item H/O liver;
\item HLA match;
\item gender mismatch - M:M, F:F, M:F, F:M;
\item ABO mismatch;
\item CMV status;
\item pre-transplant medication including Myelotarg, Cytarabine, 6-Mercaptopurine, Azathioprine, Oxaliplatin.
\end{itemize}  
We construct the prospective patient's DT, defined with the real patient's pre-transplant variable values, where such variables fall into the categories listed above. Figure~\ref{fig:dt} is a cartoon representing such a DT. We perform the automated prediction of the VOD severity score of this DT within Model~1, and such a predicted score is then fed back to the real patient as a score of the severity with which VOD will develop in this patient, post-BMT.

\section*{Results}
\begin{figure}[ht]
\centering
\includegraphics[width=.3\linewidth]{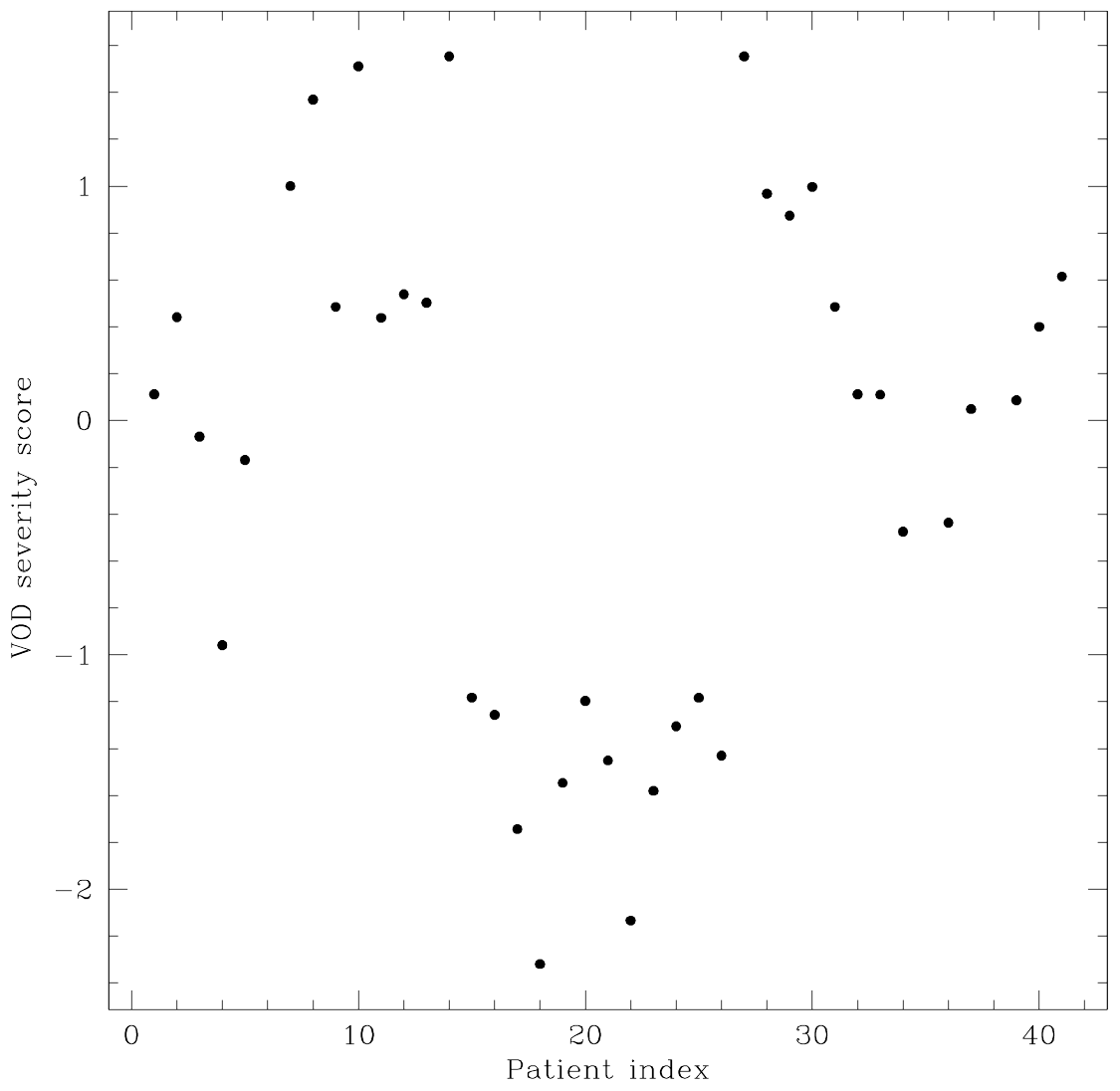}
\caption{Score that parametrises severity with which VOD develops in a patient, after they have undergone their BMT, is plotted against the index of the retrospective patient, whose information comprises a datum of the training dataset ${\bf D}_{aug}$.}
\label{fig:daug}
\end{figure}

In Figure~\ref{fig:daug}, the 38 scores included within augmented training set ${\bf D}_{aug}$ are plotted against the patient index.

In Table~\ref{tab:1}, we present the mean score and the 95$\%$ Highest Probability density credible regions (HPDs) on the score, learnt (using Model~2) for the eight prospective patients, post-\cite{plos}.
\begin{table}[ht]
\centering
\begin{tabular}{|l|l|l|}
\hline
Patient index & Mean score & 95$\%$ HPD \\
\hline
ID~34 & -0.4746 & [-0.5276,-0.4260] \\		
\hline
ID~35 & -0.1357 & [-0.2501. -.0500] \\ 
\hline 
ID~36 & -0.4371 & [-0.4654,-0.4085] \\          
\hline
ID~37 & 0.0483 & [-0.0104, 0.0951] \\
\hline
ID~38 & 0.3605 & [0.2104, 0.5617] \\
\hline
ID~39 & 0.3994 & [0.2349, 0.6051] \\
\hline
ID~40 & 0.0861 & [0.0659, 0.1000] \\
\hline
ID~41 & 0.6164 & [0.5413, 0.7221] \\
\hline
\end{tabular}
\caption{\label{tab:1} Table presenting the mean and 95$\%$ Highest Probability density credible region (or HPDs) of the score learnt for the eight new patients who received BMT after the publication of \cite{plos}.}
\end{table}

\begin{figure}[ht]
\centering
\includegraphics[width=.5\linewidth]{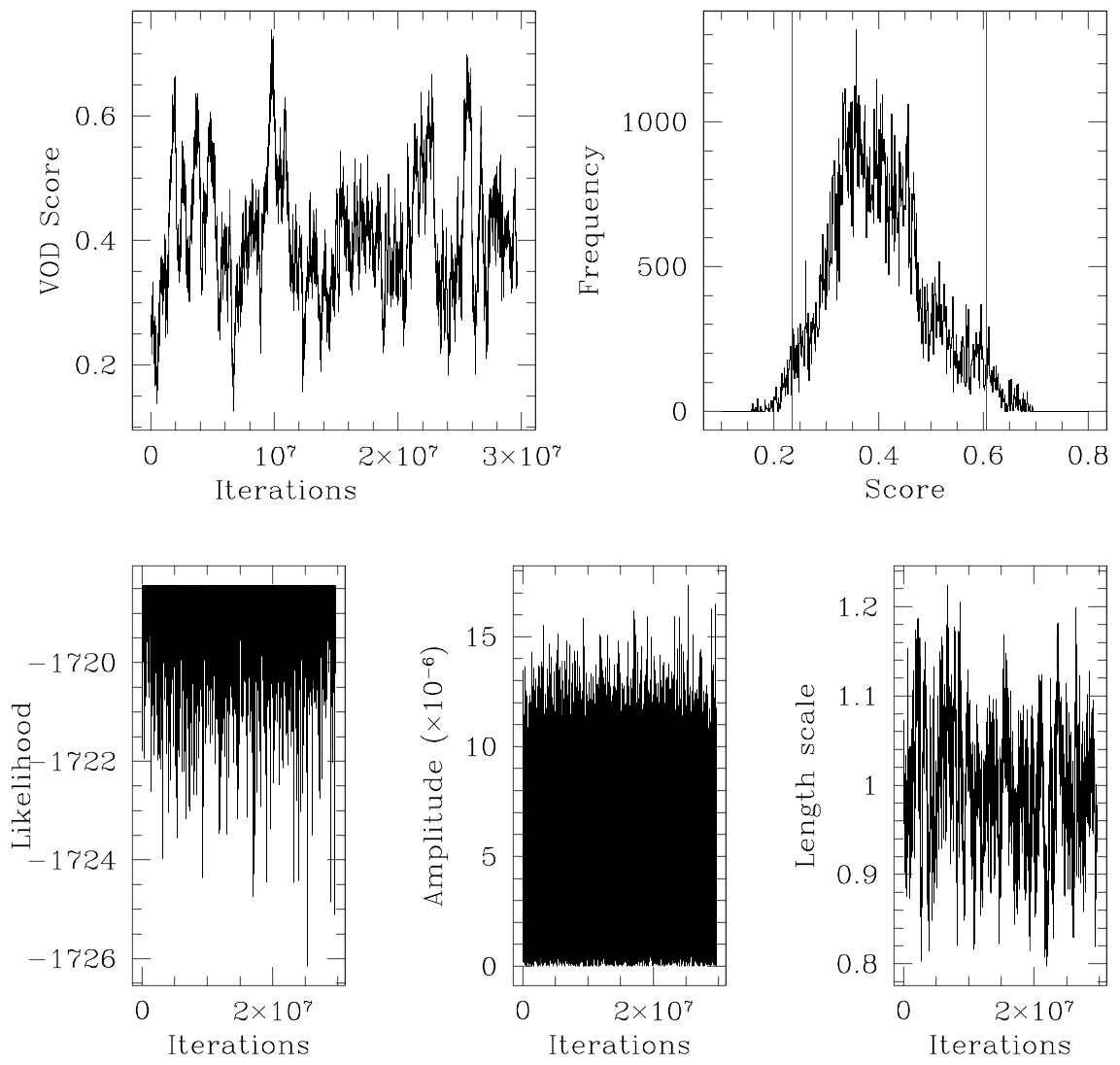}
\caption{Figure displaying results of learning undertaken for patient with ID~39. {\it{Top left:}} trace plot of the score of the severity with which VOD will develop in patient with ID~39 after their BMT. {\it{Top right:}} histogram constructed with the learnt values of the score; the vertical lines at score of 0.2450 and at score of 0.6050 indicate the left and right bounds of the 95$\%$ Highest Probability Density credible region on the learnt score.
  {\it{Bottom left}} trace plot of the learnt length scale $\ell$. {\it{Bottom middle:}} trace plot of the amplitude parameter $A$. {\it{Bottom right:}} trace plot of logarithm of the likelihood across iterations.}
\label{fig:samar}
\end{figure}

In Figure~\ref{fig:samar} we display results of performing the learning of the score of the patient with ID~39, along with the learning of the model parameters $\ell$ and $A$, given the data on all patients available at the time the score learning for this patient was undertaken. This included the data of all 25+7 patients whose scores were reported in \cite{plos}, augmented by the learnt scores of patients with ID~34, ID~36 and ID~37; recall that patients with ID~35 and ID~38 are left aside for the validation exercise with Model~1. We refer to this dataset as ${\bf D}_{39}$. Indeed, values of the pre-transplant variable vector $\bX$ of all 35 patients who are included in dataset ${\bf D}_{39}$ were known, in addition to the learnt values of their score variable. Figure~\ref{fig:samar} displays the traces of parameters learnt within the MCMC chain that is run with the training set ${\bf D}_{39}$, where by ``trace'' plot of a parameter, we mean the plot of the variation of values of the parameters noted across iterations of the MCMC chain. This patient's mean VOD score is learnt as 0.3994, and the uncertainty in the score learning is given by the 95$\%$ Highest Probability Density credible region [0.2450, 0.6050]. Trendlessness of the trace plots in Figure~\ref{fig:samar} manifest convergence attained by the MCMC chain. While results of the learning within Model~2 for patient with ID~39 are shown in this figure, it is such learning that is undertaken for all patients with ID~$k$, for $k \in\{34, 35, \ldots, 41\}$.

\subsection*{Equipping Model~1 to perform closed-form prediction}
As we have discussed above, it is highly desirable to use Model~1 to learn the function that outputs the score variable $S$ that is realised at a value of the vector $\bX$ of pre-transplant variables, so that we can subsequently perform the closed-form prediction of the mean and uncertainty of the score variable $S_{test}$ of the DT of a prospective patient, where $S_{test}$ is realised at input $\bx_{test}$ which is the value of the pre-transplant variable vector that defines this DT. This information on the DT's predicted score is then fed back to the prospective patient in question, (see Figure~\ref{fig:dt}), to enable the physicians to decide on the course of action regarding early intervention against VOD.

Learning with Model~1 demands the learning of the 30 length scale parameters $\ell_1,\ldots,\ell_{30}$ of the kernel that parametrises the correlation function of the stochastic process that generates the sought function (that outputs score at a value of the pre-transplant variables vector). Such learning is undertaken with the training dataset ${\bf D}_{aug}$.

\begin{figure}[ht]
\centering
\includegraphics[width=.8\linewidth]{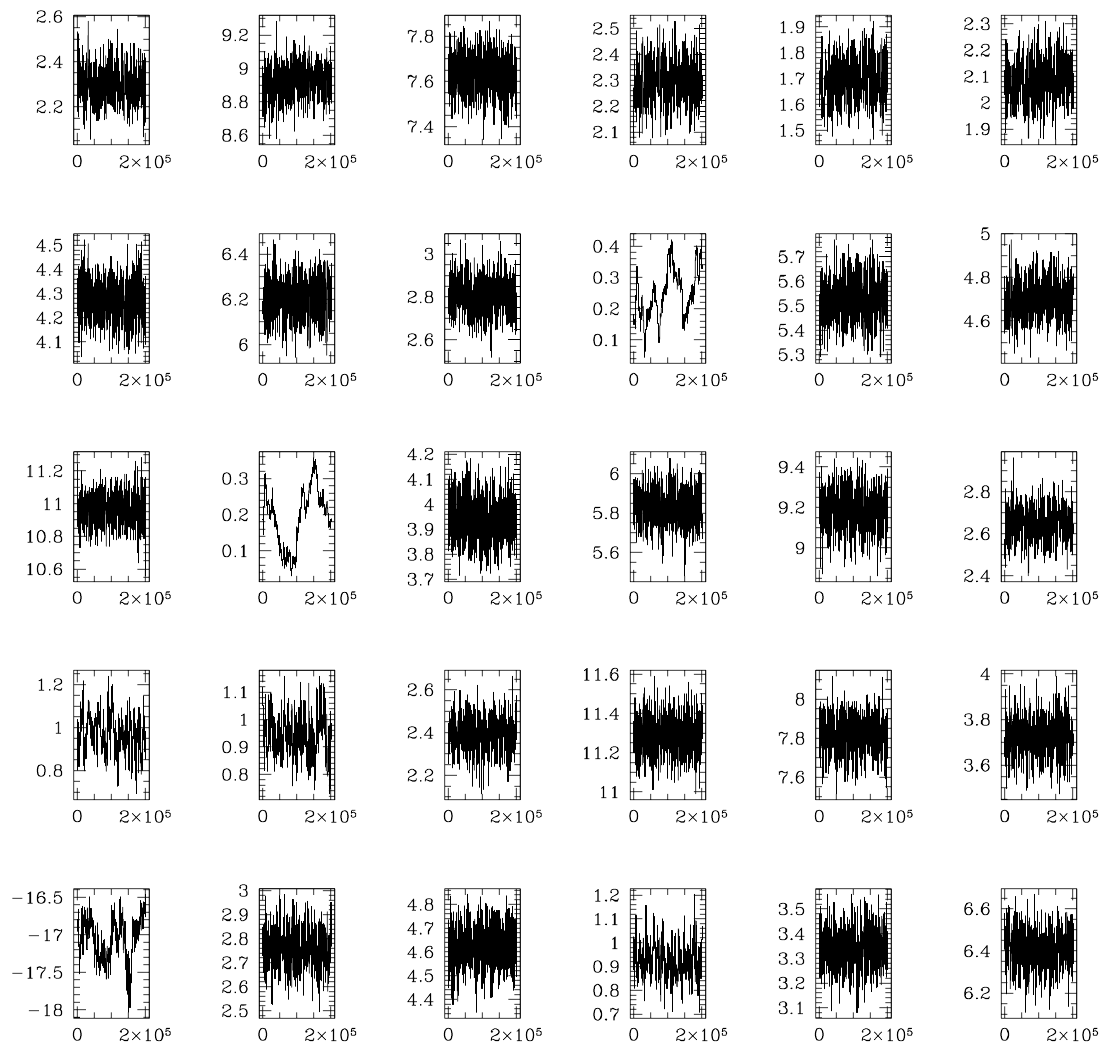}
\caption{Figure displaying trace plots of the length scale parameters $\ell_2, \ell_3, \ldots, \ell_{30}$ in all panels of the plot, except in the bottom left-most panel, in which we have plotted the logarithm of the likelihood of $\ell_1,\ell_2,\ldots, \ell_{30}$ in the data ${\bf D}_{aug}$.}
\label{fig:ells}
\end{figure}

In Figure~\ref{fig:ells}, we present the results of learning the length scale parameters of the kernel. Again, it is the trace plot of each learnt length scale parameter that we display, and monitor for its trendlessness, as an indicator of the convergence of the MCMC chain run using the data ${\bf D}_{aug}$ that has a size of 38. 

\subsection*{Closed-form score prediction}
\begin{figure}[ht]
\centering
\includegraphics[width=.5\linewidth]{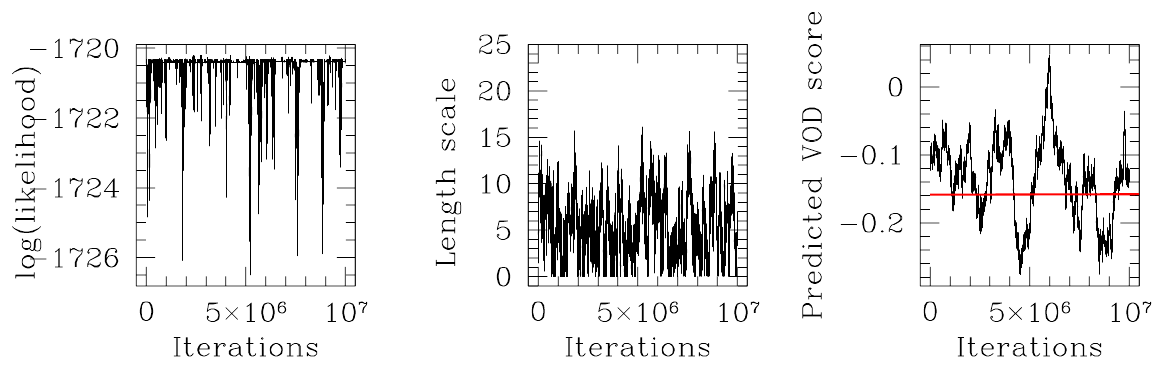}
\caption{Figure displaying trace plots of the likelihood (on the left), length scale parameter $\ell$ (in the middle panel), and the score learnt using training data ${\bf D}_{aug}$ using Model~2, at the known value $\bx_{35}$ of the vector of pre-transplant variables for patient ID~35 (on the right). In the red (or grey in the monochromatic version) horizontal line, we overplot the mean score of the DT with ID~35, predicted in a closed-form way within Model~1. }
\label{fig:35}
\end{figure}

\begin{figure}[ht]
\centering
\includegraphics[width=.5\linewidth]{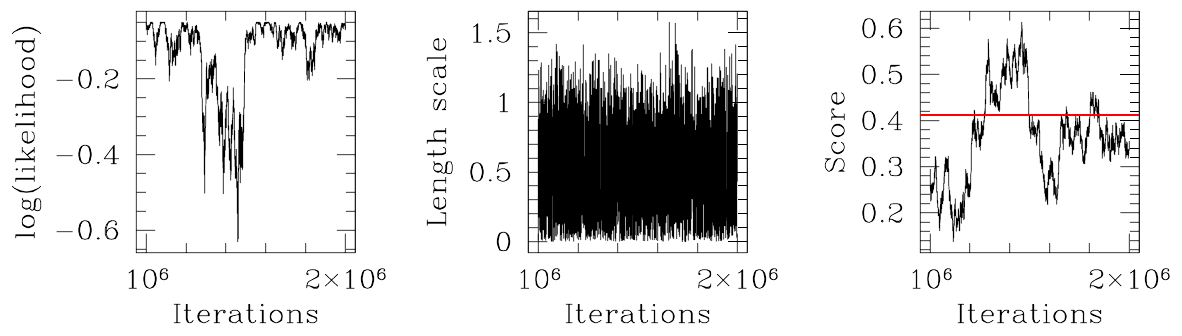}
\caption{Same as in Figure~\ref{fig:35}, except here we plot results for patient ID~38.}
\label{fig:38}
\end{figure}

Using the mean of the learnt values of the 30 length scales in Model~1, we then defined the kernel that uniquely parametrises the correlation function of the stochastic process, a sample function of which is a model of the sought function $f(\cdot)$ that outputs the severity score. Indeed, once the kernel is specified, we arrived at the position to predict the output of this learnt function, at the value of the vector of pre-transplant variables that define the DT of the prospective patient being treated. Such prediction is closed-form, and therefore automated, in addition to being interpretable and reliable. In fact, we predict the score for the DT with patient ID~35, and again for the DT with patient ID~38. Said predictions are depicted in Figure~\ref{fig:35} and Figure~\ref{fig:38} respectively. We compare the score predictions performed at the known pre-transplant variables of each of these patient DTs, with the score value that is learnt for the respective patient, using dataset ${\bf D}_{aug}$ within Model~2. We recall that using Model~2 results in the learning of the length scale $\ell$ and the amplitude $A$ as well. In each of Figure~\ref{fig:35} and Figure~\ref{fig:38}, we plot the traces of the likelihood; length scale parameter; the score variable - and overplot the mean score of the DT of this patient, predicted in a closed-form way within Model~1, upon the trace of the learnt score. The DT's mean score obtained via the closed-form prediction allowed within Model~1, is noted to be comfortably included within the uncertainties on the score learnt using Model~2, for both test cases, namely patient ID~35 and patient ID~38. 

\subsection*{Facility to perform automated pre-transplant prediction}
\begin{figure}[!ht]
\centering
\includegraphics[width=.75\linewidth]{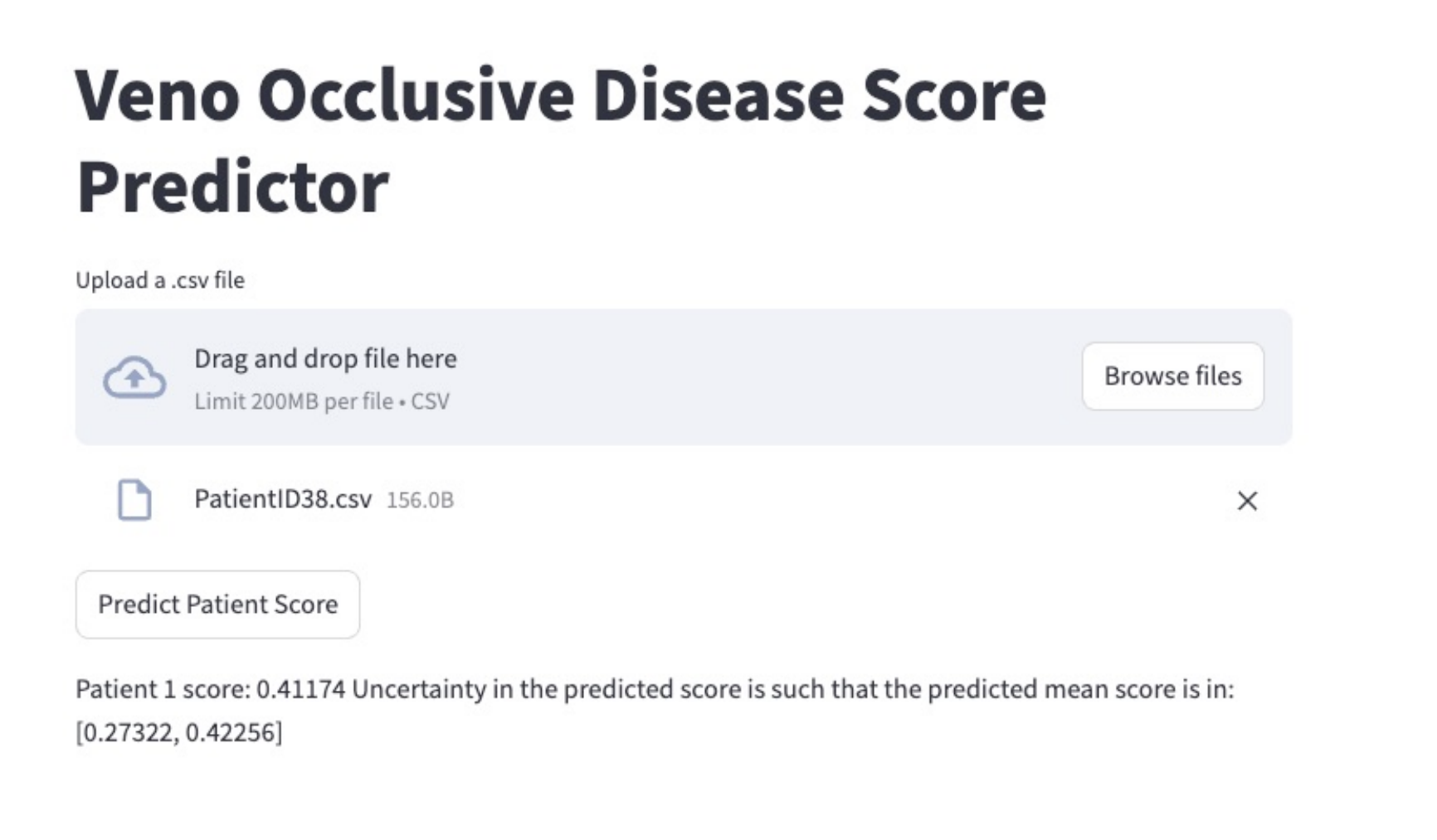}
\caption{Figure displaying a snapshot of a web-based application developed to undertake early (i.e. at the pre-transplant stage), individualised prediction of the score for the DT of a considered prospective patient. The app accepts the pre-transplant variable values that define this DT, as a csv file. Upon receipt of this file, the mean score and the uncertainty in the score are then predicted. The app is password-secured.}
\label{fig:app}
\end{figure}

Identification of the 30 length scale parameters of the kernel allows for the automated pre-transplant prediction of the severity score of the DT of prospective patient, which in turn informs on the score of the severity with which VOD will develop in this prospective patient after they have undergone their BMT. We have developed a web-based (password-guarded) application that performs such automated prediction. Figure~\ref{fig:app} offers a snapshot of the said app. The app requires a csv file containing values of the pre-transplant variables of the DT, and it outputs the predicted mean severity score, along with the predicted uncertainty on this score. Such an automated predictive facility is very helpful for the Haematologist-Oncologists in deciding on whether to start the relevant prospective patient on a course of Defibrotide, before their BMT has started.

\subsection*{Relevance of predicted score in VOD treatment}
In our team, if the predicted VOD severity score is high or if the clinical suspicion for VOD onset is high - based on the Baltimore criterion, and radiological inputs such as the Hokus scoring on abdominal USG - the treating physicians recommend placing the considered patient on Defibrotide from Day~-2. However, a reality they face is that Defibrotide is not available in India and needs to be imported, making it phenomenally expensive, sometimes costing in excess of the transplant itself. So in spite of their intent and recommendations, Defibrotide usage is not always possible. The doctors use prophylaxis with Heparin, Ursodeoxycholic acid, and N-Acetyl-L-Cystiene in all prospective BMT recipients.

\section*{Methods}
We seek the function $f(\cdot)$, where $S= f(\bx)$. Here $S\in{\mathbb R}$ is the random variable that represents the score of the severity with which VOD develops in a patient with pre-transplant variables $X_1,X_2, \ldots,X_p$, with the $p=30$-dimensional vector of pre-transplant variables defined as $\bX = (X_1,X_2,\ldots,X_p)^T$, where $\bX$ attains a value $\bx$ in a considered patient. In the Bayesian approach that we espouse, we treat the unknown as a random variable, i.e. the sought function $f(\cdot)$ is treated as a function-valued random variable. Since the probability distribution on the space of functions is given by a stochastic process, we treat $f(\cdot)$ as a random sample function of a stochastic process. We choose this process to be a GP (that we denote $GP$); reasons for this choice are stated in the ``Background'' section.

Thus, in our model, $f(\cdot) \sim GP(\mu(\cdot), \kappa(\cdot,\cdot; \bell))$, where $\mu(\bx)$ is the mean function of the process $GP$ computed at $\bX=\bx$, and $\kappa(\bx_i, \bx_j; \bell)$ is covariance function of $GP$, computed at the input values $\bx_i$ and $\bx_j$, such that this covariance function bears $d$ parameters $\ell_1,\ldots,\ell_d$, with $\bell=(\ell_1,\ldots,\ell_d)^T$. For us, the minimal number of parameters that we need to learn is $d=p (=30)$. Thus, $Cov(S_i, S_j) = \kappa(\bx_i, \bx_j; \bell)$. Then $GP$ induces a likelihood (of $\ell_1,\ldots,\ell_d$ in the available training data), that is the multivariate Normal density with a mean vector $\bmu$ and covariance matrix $\bSigma$. As an aside, we note that the kernel parameters $\{\ell_i\}_{i=1}^d$ are typically referred in the literature as ``hyperparameters'' of the kernel that parametrises the covariance function of the considered Gaussian Process \cite{rasmussen}; however, here we refer to $\ell_i$ as the $i$-th ``parameter'' of the kernel, $\forall i\in\{1,\ldots, d\}$.

We standardise the sample $\{s_1, s_2, \ldots, s_N\}$ of known values of the output score variables of $N$ retrospective patients, realised at the design points $\bx_1, \bx_2, \ldots, \bx_N$ that are the known values of the vectors of pre-transplant variables of these patients. Here, the output score variable $S_i=f(\bx_i)$ is realised as $s_i$, for any $i\in\{1,\ldots,N\}$. Then $\bmu$ is rendered a null vector while $\bSigma=[Cov(S_i,S_j)]$ is rendered a correlation matrix. Then a sample function $f(\cdot)$ of $GP$ can be generated by learning the signature correlation function of $GP$, i.e. by learning the kernel parameters $\ell_1,\ldots, \ell_d$ given the training data ${\bf D} = \{(\bx_i, s_i)\}_{i=1}^N$. Then setting $\bs=(s_1,\ldots,s_N)^T$, in Model~1, the multivariate Normal likelihood of $\ell_1,\ldots,\ell_d$ in ${\bf D}$ is denoted ${\cal MN}(\bmu, \bSigma)$, and written out clearly as:
\begin{equation}
{\cal L}(\ell_1,\ldots,\ell_d; {\bf D}) = \displaystyle{\frac{1}{\sqrt{\vert
      2\pi\bSigma\vert}} \exp\left(-\frac{(\bs - \bmu)^T \bSigma^{-1}
      (\bs-\bmu)}{2}\right)}.
 \label{eqn:likeli}
\end{equation}

\subsection*{Kernel parametrisation of $\bSigma$ given mixed-type input variables}
Here we parametrise the $i,j$-th element $Corr(S_i,S_j)$ of $\bSigma$, using the kernel $K(\bx_i,\bx_j; \ell_1,\ldots, \ell_d)$, for $d=p$, where the input vector $\bX$ is $p$=30 dimensional; $\forall i,j\in\{1,\ldots,N\}$. Given the small size of the dataset that is available to us, we invoke a simple stationary SQuare Exponential (or SQE) kernel.

In a generic application with a $q$-dimensional vector $\bY$ as the input variable, this SQE kernel has the following form, when computed at the input values of $\by$ and $\by^{\prime}$: $\exp(-(\by - \bmu)^T \bPsi (\by - \bmu)/2)$, with $\bPsi$ a $q\times q$-dimensional diagonal matrix, such that the $c$-th diagonal element is $1/z_c^2$. The interpretation of $z_c$ is that it is the length scale along the $c$-th direction of input space, for $c\in\{1,\ldots,q\}$.

Unlike the generic example above, the input vector $\bX=(X_1,\ldots, X_p)^T$ in our work is of mixed-type, i.e. some components of $\bX$ are numerical - namely, $X_{14}, X_{16}, X_{17}, X_{19}$ - while all other components are binary. Given this, we adopt the SQE kernel as follows: 
\begin{equation}
  K(\bx_i, \bx_j; \ell_1,\ldots, \ell_d) =
  \displaystyle{\exp\left(-{\sum\limits_{c\in {\bN}} \frac{(x_i^{(c)} - x_j^{(c)})^2}{2\ell_c^2}} - {\sum\limits_{c\notin {\bN}} \frac{(1 -\delta( x_i^{(c)}, x_j^{(c)}))^2}{2\ell_c^2}}\right)},
  \label{eqn:kernel}
  \end{equation}
where the delta function $\delta(a,b) = 1$, if $a=b$ and $\delta(a,b) = 0$, if $a \neq b$, while the set $\bN=\{14, 16, 17, 19\}$ bears the indices of the numerical components of the input vector. This holds for all $i,j\in\{1,\ldots,N\}$. This kernel was introduced by \cite{hutter}, and we had formulated it independently for the learning using Model~1.

Thus, by recalling that $Corr(S_i,S_j) = K(\bx_i, \bx_j; \ell_1,\ldots, \ell_d)$, we can define $\bSigma = [K(\bx_i, \bx_j; \ell_1,\ldots, \ell_d)]$, where this kernel is given in Equation~\ref{eqn:kernel}. Thus $\bSigma$ is defined and this in return allows for the likelihood to be computed in Equation~\ref{eqn:likeli}.

We recall that $N=38$ in our work with Model~1.

\subsection*{Inference}
We thereafter invoke adequate priors $\pi_0(\ell_1,\ldots,\ell_d)$ on $\ell_1, \ldots, \ell_d$, to formulate the joint posterior probability density of the $d$ unknowns, given the data ${\bf D}$ as $\pi(\ell_1,\ldots,\ell_d\vert{\bf D})\propto \pi_0(\ell_1,\ldots,\ell_d){\cal MN}(\bmu,\bSigma)$. We generated samples of each sought parameter, from this joint posterior density, using MCMC techniques.

Given that $\ell_i\in{\mathbb R}$, during the $t$-th iteration of the undertaken chain, we propose $\ell_i$ from a Normal proposal density as: $\ell_i^{(*t)}\sim {\cal N}(\ell_i^{(t-1)}, \sigma_i^2)$, where $\ell_i^{(t)}$ is the current value of the $i$-th length scale parameter $\ell_i$, in the $t$-th iteration. Then given that the MCMC chain is $N_{iter}$ iterations long, $t\in\{1,\ldots, N_{iter}\}$, with $i\in\{1, \ldots, d=p=30\}$. Thus, we use the Random Walk Metropolis Hastings algorithm for the MCMC. Here, for each $i$, we choose the jump scale $\sigma_i$ as a small fraction, (typically $0.1\%$ to $0.5\%$), of the seed value $\ell_i^{(0)}$ of $\ell_i$, i.e. the value of $\ell_i$ used to start the MCMC chain. We also do not have strong priors on $\ell_i$, for any $i$, or on any function of $\ell_1,\ldots,\ell_d$. Given this, we use weak priors that we choose to be $\pi_0(\ell_i) = {\cal N}(\ell_i^{(0)}, \sigma_{i,0}^2)$, where the prior variance $\sigma_{i,0}^2$ is chosen to be $> 10$ times $\ell_i^{(0)}$.

In fact, such a choice characterises the ``pilot chain'', that we run for a large $N_{iter}$ with a seed value of 1, and the aforesaid weak priors, for all length scale parameters. We experiment with the jump scales until a semblance of convergence is achieved, as tracked by the trace plots of $\ell_1,\ldots \ell_d$, as well as the trace of the logarithm of likelihood. Once this pilot chain has converged, we note the mean of the post-burnin values of each length scale parameter, and start a fresh MCMC chain, at these mean values. We also note the estimated variance of the post-burnin samples generated for $\ell_i$ in the pilot chain, and use this as the prior variance of $\ell_i$ in the fresh chain that we run; this is undertaken for all $i\in\{1,\ldots,d\}$. In this way, we achieve the results that are displayed in Figure~\ref{fig:ells}. In our work, $N=38$.

The predicted expectation of the output variable $S_{test}$ that is realised at the test input $\bX=\bx_{test}$ that defines the DT of the prospective patient under consideration, is then ${\mathbb E}(S_{test}) = \bSigma_{test}^T \bSigma^{-1}\bs$ and $Var(S_{test})=1 - \bSigma_{test}^T\bSigma^{-1}\bSigma_{test}$ where $\bs=(s_1, s_2, \ldots, s_N)^T$ and $\bSigma_{test}$ is the $p$-dimensional vector, the $i$-th component of which is $Corr(S_{test}, S_i)$, where $i\in\{1,\ldots,p\}$. Upon prediction, we unstandardise the predicted mean score and the predicted variance, using the sample mean and variance estimated for the sample $\{s_1,\ldots,s_N\}$, that we had standardised all scores by, before starting the learning.   

The inference of 30 parameters is a non-trivial exercise, and we have performed extensive experimentation to learn the length scale parameters.

\subsection*{Learning VOD score with Model~2}
In Model~2, we set $\bX$ as the output and $S$ as the input to a vector-valued function $\bog(\cdot)$, such that $\bX = \bog(s)$, i.e. $(X_1, X_2, \ldots, X_p)^T = (g_1(s), g_2(s), \ldots, g_d(s))^T$, where $g_i(\cdot)$ is the $i$-th component of the vector-valued function $\bog(\cdot)$, $i\in\{1,\ldots,p=30\}$. Then to learn $\bog(\cdot)$, we model $\bog(\cdot)$ as a random function that is treated as a sample function from a vector-valued GP that we denote ${\bf GP}$. Then by definition of a Gaussian Process, the joint probability density of $M$ realisations of a sample function drawn from ${\bf GP}$ is the higher-dimensional generalisation of the multivariate Normal density - to be precise, the matrix Normal density. Then the likelihood is a matrix Normal density, parametrised by a mean matrix; an $M\times M$-dimensional inter-realisation covariance matrix $\bSigma_{M}$; and a $p\times p$-dimensional inter-component covariance matrix $\bSigma_p$. We use a zero-mean GP, so that the mean matrix is a null matrix. Also, the $i,j$-th element of $\bSigma_M$ is $Cov(\bX_i, \bX_j)$, which we parametrise with a kernel $K(s_i, s_j; \ell, a)$ that is defined such that as the difference between $s_i$ and $s_j$ increases, the covariance between the outputs $\bX_i$ and $\bX_j$ (that are respectively realised at $s_i$ and $s_j$) decreases. Here $\ell$ is the scale length over which covariance between the outputs decreases by a stipulated factor, and $a$ is the amplitude of the covariance. In this model, we use global values of $\ell$ and $a$, i.e. the kernel hyperparameters are considered to be independent of $i$ and $j$, $\forall i,j\in\{1,\ldots,M\}$. Thus, $K(\cdot,\cdot; \ell,a)$ decreases with $\vert s_i - s_j\vert$. We use a simple form of this kernel, namely the SQE discussed above, so that $Cov(\bX_i, \bX_j) = a\exp(-(s_i-s_j)^2/\ell^2)$. We learn $\ell$ and $a$ using the training data ${\bf D}_2 = \{(\bx_i,s_i)\}_{i=1}^M$. Also, we estimate the $q,r$-th element $Cov(X_q, X_r)$ of $\bSigma_p$ using the sample comprising recorded values of $X_q$ and $X_r$, where $q.r\in\{1,\ldots,p\}$. Thus the likelihood of $\ell$ and $a$ in the data ${\bf D}_2$ is the matrix Normal density denoted ${\cal MT}(\bzero,\bSigma_p, \bSigma_M)$. Then using (weak) prior $\pi_0(\ell, a)$ on $\ell$ and $a$, we define the joint posterior probability density $\pi(\ell,a\vert {\bf D}_2)$.

Then for the test data point for which we know the value $\bx_{test}$
of the output $\bX_{test}$ - realised at $S=s_{test}$ where $s_{test}$
is not known - we want to compute the joint probability density of
$S_{test}, \ell, a$. This joint is $\pi(S_{test}=s_{test}, \ell, a \vert {\bf D}_2, \bx_{test})$, which is given by
$\pi(\ell, a \vert {\bf D}_2, \bx_{test}, s_{test}) \times
f({\bf D}_2, \bx_{test}, s_{test})$. While
the first factor in the right hand side of the last equation is known
as ${\cal MT}(\bzero, \bSigma_p, \bSigma_{M+1})$, the second factor
is expressed using the Law of Total Probability as:
$f(s_{test}, \bx_{test}, {\bf D}_2) =
\int_{\ell\in{\mathbb R}}
\int_{a\in{\mathbb R}_{> 0}}
f(\bx_{test}\vert {\bf D}_2, s_{test}, \ell, a)
f({\bf D}_2, \bs_{test}, \ell, a) d\ell da$.
The second factor of the
integrand on the right hand side is a prior density (on $\bs_{test}, \ell, a$), that we choose, while the first factor is the known closed-form posterior
predictive of the output variable realised at $S=s_{test}$, given
chosen values of the kernel parameters $\ell$ and $a$, and the available data. 
We choose (a Normal prior on $a$ and) the prior on $\ell$ to be Uniform over a chosen interval $[-\ell_0,\ell_0]$, and the prior of $S_{test}$ to be the integral of the known posterior predictive of $\bX=\bx_{test}$, at $S=s_{test}$ - given ${\bf D}_2, \ell, a$ - with respect to $\ell\in[-\ell_0, \ell_0]$. Then we get that
$\pi(S_{test}, \ell, a\vert {\bf D}_2,
\bx_{test}) \propto \pi(\ell, a\vert {\bf D}_2, s_{test},
\bx_{test})$

is a global constant times the matrix Normal density parametrised by a null matrix, $\bSigma_p$ that we estimate, and $\bSigma_{M+1}$ that we learn. Since MCMC allows learning, irrespective of global constants,
we generate samples comprising
values of $S_{test}, \ell, a$, given the data ${\bf D}_2$ and the
known $\bx_{test}$.

So in the $t$-th iteration of the MCMC chain, we propose $s_{test}^{(\star,t)}$ from a Normal density with mean $s_{test}^{(t-1)}$ and variance $\sigma_{test}^2$ that is set as an experimentally-chosen (small) fraction of $s_{test}^{(0)}$. Again, $\ell$ is proposed from a Normal density centred at the current value of $\ell$ and a distinct, experimentally-chosen variance. $a$ is proposed from a truncated Normal density (truncated on the left at 0), with mean given by the current value of $a$ and a distinct, experimentally-chosen variance. We also use the priors stated above on these three parameters, during the ``pilot chains''. In these pilot chains we set $s_{test}^{(0)}=0$; $\ell^{(0)} = a^{(0)} = 1$. Once a semblance of convergence is achieved, we start a final chain in which we use the mean of the post-burnin samples as the seeds, and the same priors as in the pilot chain. The inferred values of $S_{test}, \ell, a$ are presented as the respective 95$\%$ Highest Probability density credible region, accompanied by the mean of this credible region. 

\section*{Conclusions}
In this paper we have discussed a new approach that facilitates individualised prediction of the severity with which a disease will develop in a patient, where such a prediction is performed at the pre-onset stage, i.e. before the symptoms of the disease show up clinically. We have illustrated this approach to predict the severity score of VOD in the DT of a real prospective patient, where the DT is defined using information on the pre-transplant state of this prospective patient. Thus, our prediction is made at the pre-transplant stage, and it tells the treating physicians - at this early stage - how severely VOD will develop in this patient after they would undergo a BMT. This then allows the physicians to undertake early intervention (for example, with Defibrotide), to prevent development of VOD, if such intervention is relevant for the patient at hand, as judged at the pre-transplant stage, based on the VOD severity score predicted for this patient's DT. We have wrapped this automated predictive capacity within an AI facility so that the Haematologist-Oncologists can undertake the automated prediction independently of their probabilist collaborators, by inputting all parameters that in our model, inform on the pre-transplant state of the prospective patient at hand, i.e. on this patient's DT.

Continuing the prediction of the severity score of further patients will allow for the updating of the training dataset that we employ, to learn the parameters of (the kernel that parametrises the correlation function of the) stochastic process that generates the sought pre-transplant variables-to-severity score relationship. Indeed, quality and reliability of the predictions will improve with the increase in size of, and in the variety within, the training set. Appending the predictions reported here, to the existing training set, augments the data size to 40, with which we will undertake few more predictions, before we will update our learning of the process parameters. We will continue to update the learning of the process parameters after every five or fewer new predictions.

Beyond this work, our interest is in adopting this approach for the early individualised prediction of the severity score of other diseases in prospective patients. In particular, we are keen on doing this for Cytokine Release Syndrome (or CRS). This will require the learning of a random graph given the time series data on CRS phenotypic markers that evolve with time, as observed during the time interval that includes the day the BMT in performed for a patient, where such parameters will include binary variables noted in the patient charts - such as the ``incidence of a rash'' parameter that might attain a value of 1, the $t$-th day onward, (indicating a rash in the patient, the $t$-th day onward), while this parameter is 0 for all times before $t$. Other temporally-evolving phenotypic markers are numerical. We will undertake the learning of the correlation between pairs of binary variables using the Cramer's V measure of correlation, while the correlation between a numerical and binary variable can be computed using the point-biserial measure. Correlation between numerical variables will be computed using the Spearman correlation measure. Once the graph is learnt, a statistical distance between the graph variables of the patient datasets (borne in their respective charts) will be employed, and post-BMT specifics related to CRS onset acknowledged, to compute the CRS severity score of each patient in a retrospective cohort, relative to that of an arbitrarily-chosen reference patient. This will then allow the originally-absent training set to be populated with pairs of values of pre-transplant state vector and CRS score, of patients in this cohort, to finally permit prediction of the CRS score of a prospective patient, at the pre-transplant stage.      

\bibliography{refs}

\section*{Acknowledgements}

KW acknowledges an internal Impact fund; CZ acknowledges funding under the project UKRI2398.

\section*{Funding}
There is no funding to report for this work.

\section*{Author contributions statement}

{DC, KW, CZ contributed to the learning and prediction using the data produced by AB, JC. All authors contributed to paper writing and editing}

\section*{Additional information}
We have included the code to learn the function that outputs the score variable, at a given value of the vector of pre-transplant variables, and performs prediction of the score for a test patient.

\end{document}